\documentclass[12pt,a4]{article}
\def\beq{\begin{equation}}
\def\eeq{\end{equation}}
\usepackage[dvips]{graphicx}
\begin{document}
\title{A Graph method for mapping changes in temporal and spatial
phenomena with relativistic consequences}
\author{Daniel Brown}
\newcommand{\ictitle}
{A Graph method for mapping changes in temporal and spatial
phenomena with relativistic consequences}

\newcommand{\icauthor}
{Daniel Brown\footnote{email: d.brown@cs.ucl.ac.uk}\\Depts of
Physics and Astronomy and Computer Science}
\newcommand{\icaddress}
{University College London\\ Gower Street London WC1E 6BT, U.K.}

\begin{titlepage}
\begin{center}
{\large{\bf\ictitle}}
\bigskip \\ \icauthor \\ \mbox{} {\it \icaddress} \\ \today \\ \vspace{.5in}
\end{center}
\setcounter{page}{0}
\begin{abstract}
A cellular automata approach (using a Directed Cyclic Graph) is
used to model interrelationships of fluctuating time, state and
space. This model predicts phenomena including a constant and
maximum speed at which any moving entity can travel, time dilation
effects in accordance with special relativity, relativistic
Doppler effects, propagation in three spatial dimensions, an
explanation for the non-local feature of collapse and a
speculation on an explanation for gravitation effects. The
approach
has proven amenable to computer modelling. \\

A further paper details the statistical implications
for identifying the probability of locating a particle at a
particular position in
space.\\

\end{abstract}
\end{titlepage}

\section{Introduction}

Minsky (1982) investigated a model of the universe using ``a
crystalline world of tiny, discrete `cells', each knowing only
what its neighbours do". In Minsky's model properties such as a
maximal speed emerge. However, Minsky found that the model rapidly
lost coherence requiring increasing additional rules and the wrong
time dilation factors emerged. Feynman (1982) also examined
cellular automata models and was particularly concerned with
simulating time on computers using a model of discrete time; he
noted that ``a very interesting problem is the origin of the
probabilities in quantum mechanics". Recent research, such as
Jaroskiewicz (2000) has resurrected analysis of cellular automata
using an approach centred on the evaluation of non-local
information. The current paper addresses these issues through a
cellular automata method using several dimensions of time. Whilst
Tegmark (1997) considered that 3 dimensions of space with more
than one dimension of time produces ``unpredictable" artifacts
such as backward causation, this paper aims to demonstrate that
this is not necessarily the case if the time dimensions are
appropriately formulated.

\section{Background}

Explanations for a number of physical phenomena remain
unexplained. These include wave/particle ``duality'', the reason
for a maximum possible speed and ``action at a distance'' effects.
This paper resolves these phenomena through an analysis of the
nature of \emph{changes in time}.

To model these changes, the approach has two features of
particular note: it is distributed and logical precedence has
priority over all other conditions (including temporal
precedence). It uses directed (cyclic) graphs; Pearl notes this is
an excellent apparatus for study since ``causality has been
mathematicised" (Pearl 2000) - and they lead naturally to Markov
modelling.

A graph comprises a collection of entities (or nodes or vertices)
connected together by links (edges). The value of any entity can
be measured, but to predict its value the values of other
interrelated entities and the rules for their combination have to
be known also.\\

An Entity is defined through four principal components: its
\emph{elements}, the \emph{rules} which govern the cycle between
these elements, the \emph{links (or triggers)} that initiate
cycling between elements, and the \emph{constellation} which maps
the links to other entities in the graph.

A constellation of linked changing entities establishes an
Interrelated Fluctuating Entity (IFE) disturbance.

Each Entity contains a set of elements e.g. (0,1,2,3,4,5). An
entity has a minimum of 2 elements and no maximum. Once triggered
(by a link from another entity) the entity can be set to cycle
through its
sequence of elements as follows:\\
 (i) cycle forward a single element only until a further trigger\\
  (ii) cycle through the complete set of
 elements\\
 (ii) cycle backward through the set of elements\\
(iii) On reaching a specified element value (e.g. 5) the entity
can be determined to:
\\  \hspace* {10mm}(a) return directly to the first element \\\hspace* {10mm}(b)
cycle in steps back to the first element (1,2,3,4,5,4,3,2,1) \\
\hspace* {10mm}(c) remain at the specified element value

A graph \textit{links} two or more separate entities. An entity
starts cycling through its elements when triggered by a specified
change in a linked entity. All links between IFEs are directed (a
trigger by one entity logically activates the cycling of another
entity) and can be cyclic (e.g. where an element in X triggers an
element in Y
and an element in Y triggers an element in X). An entity can be determined to trigger an adjacent entity: \\
(i) By any change in element value
\\ (ii) By passing a specific element value\\
An important link is via a ``trigger'' threshold value p. Thus the
entity Space=(1,2,3,4,5) can be set to trigger the entity
Time=(1,2,3,4,5,6,7,8,9) when Space reaches the
Space$\rightarrow$Time-change trigger value p=4.\\

\textbf{Example: how to establish a wave/particle disturbance} \\
A graph can model a moving disturbance. A simple example of this
is a row of football fans creating a ``Mexican Wave'' in a stadium
(similar to a model for a series of falling dominoes). We can
model the fans using a wave pattern through a sine function, but
for the fans themselves it is easier to use a set of simple rules
such as the following... If the first fan starts to stand up, then
once this fan reaches a certain height this triggers the spatially
adjacent fan to start standing, which triggers the next fan...

Likewise, with five dominoes laid out in a one dimensional row, if
we tip the first one over to the right then the next to its right
will fall, which triggers the next one to fall...

The entity ``State'' is defined as $R=\{0,h,2h,3h\}$ where 0
indicates an upright domino, h indicates a tipping domino, 2h
indicates a domino tipping further and 3h indicates a horizontal
fallen domino.

The entity ``Space'' is defined as $x=\{0,1,2,3,4\}$ where 0
indicates the first spatial position, 1 the next spatial position
to the right etc...Each of the 5 spatial positions is therefore an
equal distance dx=1 units apart. Because State and Space have
finite numbers of elements, then by applying rule 2.(iii)(c),
entities will eventually remain fixed on their final elements.

The entity ``Time'' is defined as the infinite set of elements T =
$\{0,1,2,3,4,...\}$

The Space, Space and Time entities interrelate in the domino
graph: ((R),x,T). - The extra bracket for R indicates a distinct
State \textit{entity} for each \textit{element} of Space and Time.

A directed link from State to Time is defined such that \emph{any}
change in State triggers a change in Time $d\alpha$. This is the
State$\rightarrow$Time link.

A directed link from Time to State is defined such that \emph{a
change in time of s units} triggers a change in State (i.e. ``it
takes s units of time to transition from one State to the next").
This is the Time$\rightarrow$State link

A directed link from State to Space is defined such that a change
of State \textbf{only where R'=ph} at a spatial position x
triggers a change in Time element $d\beta$ at the \emph{adjacent}
spatial position x', with the time measure but not the State
carried forward to this next spatial position\footnote{Triggering
a change in domino State at an adjacent spatial position is
equivalent to a change in Space(dx) followed by a change in
State(dR). We can theoretically dispense with the physical
structure and regard the spatial layout abstractly as itself an
IFE which interacts with the IFEs of State and Time. Both Time and
Space then form variable pointers of an array (x,T) which contain
a value of State(R). Hence a change in Time preserves a continuity
in State at the new (x,T), but a change in Space does not.}. This
is the State$\rightarrow$Space link.

To establish the time at any given point in Space and State, for
the moment it is simply assumed that $T=\alpha+\beta$; however
this is an assumption that will be dispensed with later. The graph layout is therefore:\\

State(R) $\rightarrow$ Space(x) \\
\hspace*{10mm}$\downarrow$\hspace*{0mm} $\uparrow $\hspace*{7mm} $\downarrow$\\
\hspace*{10mm}Time(T) \\

The logical rules for this algorithm, where $\rightarrow$
signifies a transition, $T^+$ indicates the adjacent successor of
T and $\supset$ indicates a \emph{logical} implication using
declarative programming, are - where $(R,x,T)$ indicates
coordinates of (State,Space,Time):

1. a change in State dR of $\{(R,x,T) \rightarrow (R^+,x,T)\}
\supset$ a change in Time dT such that $\{(R^+,x,T) \rightarrow
(R^+,x,T^+)\}$

2. a change in Time dT of $\{(R,x,T) \rightarrow
(R,x,T^+)\}\supset$ a change in State dR such that $\{(R,x,T^+)
\rightarrow (R^+,x,T^+)\}$

3. a particular change in State dR where $R^+$=h of $\{(0,x,T)
\rightarrow (h,x,T)\} \supset$ a change in Time at a spatially
adjacent adjacent IFE such that $\{(0,x^+,0)\rightarrow
(0,x^+,T^+)\}$ (where $T^+=T+dT$)\footnote{this is equivalent to
the Time entity incrementing and moving in spatial position i.e.
$(0,x,T)
\rightarrow (0,x^+,T^+)$}.\\

Selecting p=1 and s=10, and assuming an initiating trigger of $(0,0,0)\rightarrow (0,0,h)$, the disturbance therefore advances:\\

 $(0,0,0) \rightarrow (h,0,0) \rightarrow (h,0,10) \rightarrow (2h,0,10) \rightarrow (2h,0,20) \rightarrow
 (3h,0,20)$\\
 $\hspace*{33mm}\downarrow\hspace*{0mm}$\\
 $\hspace*{34mm}\rightarrow (0,1,10) \rightarrow (h,1,10) \rightarrow (h,1,20) \rightarrow
 (2h,1,20)$\\
 $\hspace*{69mm} \downarrow\hspace*{0mm}$\\
 $\hspace*{70mm} \rightarrow (0,2,20)  \rightarrow (h,2,20) $\\
 $\hspace*{104mm} \downarrow\hspace*{0mm}$\\
 $\hspace*{105mm} \rightarrow (0,3,30)  \rightarrow $\\

 Some points are worth making here.

1. We could choose to follow either the moving disturbance
advancing across space $(h,0,0)\rightarrow (0,1,10) \rightarrow
 (h,1,10) \rightarrow (0,2,20) \rightarrow (h,2,20)...$, or the disturbance advancing through States in a stationary space
position $(0,0,0) \rightarrow (h,0,0) \rightarrow (h,0,10)
\rightarrow (2h,0,10) \rightarrow (2h,0,20) \rightarrow
 (3h,0,20)$. Ambiguity arises in the identity of a disturbance since a
change in State of $0\rightarrow1$ results in both a Space change
and a Time change which results in a further change in State. The
progress of the disturbance therefore bifurcates\footnote{This is
illustrated by the celebrated paradox of the ship of Theseus. Over
a period of time in order to repair a wooden ship (Ship 1) its
planks are replaced one by one - but in addition the original
planks are taken aside and reconstituted in identical architecture
into another ship at a different location (Ship 2). Which ship is
the original ship of Theseus... - To decide, we must define our
criterion of identity: either continuity of matter over changing
space and time (Ship 2) or continuity of space over changing
matter and time (Ship 1). Both ships represent two parallel
continuities of identity. This forms a useful model as an entity
can be conceived in two alternative spatial positions at the same
time. How we regard the identity of an entity therefore affects
both what and where we presume that entity to be. In particular,
an entity can be viewed as at two different points in space at the
same time, dependent on how we have tracked and how we collapse
its identity}.

2. The sequence of the algorithm steps is important. A change in
Time associated with a change in State occurs logically prior to a
change in Time due to a change in Space. However, temporally both
time changes occur in parallel ``at the same time".

3. Time is constructed as an entity which moves independently,
``ahead" of the State change.

4. The precedence of logical change over temporal change permits
both $(R,x,T)$ and $(R^+,x,T)$. So 2 distinct states (momentarily)
coexist at the same spatial and temporal position, logically prior
to the logically subsequent temporal transition.

5. At a given time e.g. 10 there are in fact a total of 4 states
associated with the disturbance located over 2 spatial positions.
These are: (h,0,10), (2h,0,10), (0,1,10), (h,1,10. For a given
time 20, the State and Space positions become even more uncertain.

6. Each domino has an associated local spatial Time(T) - which
cycles in tandem with State changes even after the IFE disturbance
has moved on to the next spatial position - as each domino falls
to its final horizontal state. Thus the last domino will register
a time of 60 units.

7. Time can advance in variable quantities e.g. $(0,2,0)
\rightarrow (0,2,20)$

8. Increasing the State value required to trigger a Time change at
an adjacent Space element slows down the progress of the
disturbance. Thus if p is the State$\rightarrow$Space trigger then
changing p from 10 to 20 units will slow down the speed of the
disturbance.

9. Note that for a given time there is some degree of Spatial
localisation (e.g. for Time of 20 units the disturbance could only
be found to be located at Spatial position 0,1 or 2).

\section{Analysis of matter disturbance}

It is often matter rather than its spatial position or the time
associated with that matter that is defined to constitute the
identity of a thing. However, at the microscopic level, a
different viewpoint is required.

A hypothetical subatomic particle (a theoretical unit particle
without sub-components) can be defined in terms of its Energy e,
Space position x, and Time T. It will be convenient to substitute
the concept of energy with that of State R where \emph{energy is
defined as the rate of change of State}. If State change is
quantised and the smallest unit of State change is dR, then a
variable of \emph{time} specifies the energy such that
$e=\frac{dR}{dt}$ where dt is the time taken for the State change.

An important principle can be inferred. Let the matter be observed
from one moment to the next. If nothing at all has changed in the
State of the matter\footnote{Specifically, we require a change in
the time of the viewer (which implies a change in state of the
viewer) without a change in time experienced by the matter.} it
will be assumed that time will not have progressed from the point
of view of the matter, which defines a stringent notion of
invariance. A change in time can \emph{only}
be associated with a change in State or a change in Space. The following assumptions are therefore made:\\
(i) Time, Space and State advance in quantised units\\
(ii) Time can only advance when change occurs\\
(iii) change can only occur if there is either or both:\\
 \hspace* {10mm}(a) change in State position\\
\hspace* {10mm}(b) change in Spatial position

1. There cannot \emph{logically} be a change in Time without a
change in either State or Space. Causally, for a given entity in a
specific fixed spatial position, then with no change in State
\textit{there can be no change in Time}. If an entity changes
spatial position or an entity changes State, then either of these
changes triggers an increase in Time. The earlier analysis of
fans/dominos suggests that there is an issue of identity to be
considered in a movement of State or Space.

2. These Time changes can be labelled as ``alpha-time" for changes
in State (with a unit \textbf{t{'}}) and ``beta-time" for changes
in Space (with a unit \textbf{t$^{*}$}) respectively. It will not
be assumed that these times are the same. These two times will be
kept distinct and modelled separately as $(\alpha,\beta) =
(rst',nt^*)$.

3. It will further be assumed that time can never be directly
measured and that in our (phenomenal) world only State changes are
ever measured - through which changes in time are \emph{inferred}.
Consequently we only ever measure alpha-time.

4. The combination of alpha-time and beta-time that determines
when a State change is occurring at a particular Space position.
This is central to meetings of coincident State IFEs which
determine an \emph{interaction}.

5. Because our time measurements, based only on alpha-time, may
differ from the total (alpha and beta) time, measurements of
apparently coincident events will vary. Questions immediately
follow as to the nature of these two components $(\alpha,\beta)$
and how they are resolved. A graph is therefore set up to model the movement of matter in Space and Time.\\

\textbf{Progression of an energy disturbance}

The entity ``State'' is defined as the infinite set of elements
$R=\{0,h,2h,3h...\}$ where 0 indicates a null State, h indicates
an activated State...The smallest unit of State change $dR=h$.
There are an infinite number of potential States.

The entity ``Space'' is defined as the infinite set of elements
$x=\{0,dx,2dx,3dx...\}$ where 0 indicates the first spatial
position, dx the next spatial position to the right etc...The
smallest unit of Space change is dx. Each spatial position is
therefore an equal distance dx units apart.

The entity ``alpha-time'' is defined as the infinite set of
elements $\alpha = \{0,t',2t',3t',4t',...\}$. The smallest unit of
alpha-time change is t'.

The entity ``beta-time'' is defined as the infinite set of
elements $\beta = \{0,t^*,2^*,3t^*,4t^*,...\}$. The smallest unit
of beta-time change is $t^*$.

The Space, Space, alpha-time and beta-time entities interrelate in
the matter graph: $(R,x,\alpha,\beta)$.

A directed link from State to alpha-time is defined such that
\emph{any} change in State dR triggers a change in alpha-time
$d\alpha$. This is the State$\rightarrow$alpha-time link.

A directed link from alpha-time to State is defined such that
\emph{a change in alpha-time of st' units} triggers a change in
State (i.e. ``it takes s units of alpha-time to transition from
one State to the next"). This is the alpha-time$\rightarrow$State
link

A directed link from beta-time to State is defined such that
\emph{a change in beta-time of $t^*$ units} triggers a change in
State (i.e. ``it takes a unit of beta-time to transition from one
State to the next"). This is the beta-time$\rightarrow$State link

A directed link from State to Space is defined such that a change
of State \textbf{only where R'=ph} at a spatial position x
triggers a change in beta-time element $d\beta$ at the
\emph{adjacent} spatial position x', with both time measures but
not the State measure carried forward to this next spatial
position. This is the State$\rightarrow$Space link.

These features of the graph can be summarised in the table below:\\

\hspace*{-44mm}
\begin{tabular}{|c|c|c|c|c|}

  \hline
  IFE & \textbf{State(R=rh)} & \textbf{$\alpha$Time($\alpha$= rst')}& \textbf{Space(x=ndx)} & \textbf{ $\beta$Time$(\beta =nt^*)$} \\
  \hline
  \textbf{Elements}& ($0,h,2h...\infty$) &  ($0,st',2st'...\infty$) & ($0,dx,2dx...\infty$) & ($0,t^*,2t^*...\infty$)\\
      \hline
   \textbf{Cycle Rule} & cycle one element   &cycle s elements    & cycle one element & cycle one element   \\
         &          until next trigger  &until next trigger &until next trigger
         & until next trigger \\
  \hline
   \textbf{Link/trigger}
     & d$\alpha$ or d$\beta$& dR    & State(R)$\rightarrow ph$  & dx\\

    \hline
    \textbf{Graph Constellation}\\
State(rh) $\rightarrow$ Space(nd) \\
\hspace*{-15mm}$\downarrow$ $\uparrow $\hspace*{10mm}  $\uparrow $\hspace*{10mm}$\downarrow$\\
\hspace*{-2mm}$\alpha$Time(st')~~$\beta$Time$(t^*)$~~~~~~~~~~~~ \\

\end{tabular} \\

\section{Combination of alpha-time and beta-time}

Whilst alpha and beta increments in time apply \emph{logically} in
sequence, \emph{temporally} they do not operate sequentially but
simultaneously. Since they occur from the same moment, and they
occur without reference to any external time, they occur ``at
once'' and it therefore does not make sense to simply add them
together. To establish the Time value at any given point in Space
and State, we \textit{do not} simply assert $T=\alpha+\beta$.

One approach might be to assert that the larger of the two time
components covers both time advances. This would account for their
``in parallel'' progress from the same moment, but would leave the
distinct features of the two components indiscernible. To combine
their influence, it is postulated that alpha-time and beta-time
act on distinct time \emph{axes}. A separate argument (section on
3 dimensions of space) supports this for three dimensions of
space.

\textbf{To combine these coterminous advances in time, which
proceed along different axes of alpha-time and beta-time into a
single total time, the following hypothesis is made: that as for
two axes in space these axes in time are orthogonal and hence
their combination comprises a pythagorean sum into a Time
magnitude $|T|$.}

For an IFE disturbance with a State$\rightarrow$Space-change
trigger of p (i.e. dR with $R^+= ph$) and an interval between
State changes of st', if this disturbance has moved a distance
x=ndx and at this spatial position has advanced to a State R=rh:

\beq |T| = \sqrt{(nt^*)^2+(npst'+rst')^2} \eeq

This indicates that following a series of n spatial movements, in
the final nth spatial position there follows a variable r State
movements. - Note that r may exceed the
State$\rightarrow$Space-change trigger point (i.e. $r>p$ is
possible even though it will have triggered the spatially adjacent
State). This time can be referred to as the \textit{residual state
time}.

If n is large i.e. a large distance has been travelled then the
residual state time rst' term becomes insignificant and:

$$|T| \sim \sqrt{(nt^*)^2+(npst')^2} = n\sqrt{(t^*)^2+(pst')^2}
$$

\textbf{Movement of an energy disturbance}

An IFE disturbance moving across a row can be compared with a
stationary one.\footnote{the lateral effects of entities on each
other are significant (see Appendix); however in this case we
shall simply use a lateral interaction to cross spatial
dimensions.} Row A comprises n adjacent IFE States. In row B only
two spatial positions are of concern: one at the start of the row
and the second at the nth position.

To assist visualisation of the \textit{distributed} form of the
disturbance, its propagation can be imagined as a ``Mexican Wave""
of football fans undulating in a stadium (i.e. each IFE State
represents a discrete State of a fan standing up or sitting down
\textit{in a
fixed spatial position}).\\

DIAGRAM 1 - disturbance moving in Space vs Spatially static
\\

Row A $\qquad
\triangle\triangle\triangle\triangle\triangle\triangle\triangle\triangle\triangle\triangle\triangle\triangle\triangle\triangle\triangle\rightarrow$
moving disturbance

Row B $\qquad \triangle$  position1 $\qquad \qquad \qquad \quad
~\triangle$ position n\\

The two State IFEs in Row B measure time elapsed whilst remaining
spatially stationary: their time advances by State changes only.
The disturbance in Row A travels from spatial position 1 to
spatial position n and also measures time elapsed.

Time measurements can be synchronised initially between the row A
position 1 State IFE and the row B State IFEs at Space positions 1
and n\footnote{e.g. a disturbance is initiated to move out at the
same speed left and right from the middle of row B until it
interacts with the Space positions 1 and n in row B and Space
position 1 in Row A which causes all three IFEs to start timing
and for the position 1 IFE in row A to start moving.}. When the
moving disturbance in row A is adjacent to the IFE at Space
position n in row B, these IFE States can interact and time
measurements compared - time taken to move between single rows can
be ignored if n is large.

\textbf{ Movement algorithm component}\\
An algorithm can be established for the moving particle
disturbance:

(i) Change in State(dR) of $\{(R,x,\alpha,\beta) \rightarrow
(R^+,x,\alpha,\beta)\} \supset$ a change in alpha-time(d$\alpha$)
such that $\{(R^+,x,\alpha,\beta) \rightarrow
(R^+,x,\alpha^+,\beta)\}$

(ii) Change in alpha-time(d$\alpha$) of $\{(R,x,\alpha,\beta)
\rightarrow (R,x,\alpha^+,\beta)\} \supset$ a change in State(dR)
such that $\{(R,x,\alpha^+,\beta) \rightarrow
(R^+,x,\alpha^+,\beta)\}$

(iii) A specific change in State(dR) of $\{(R,x,\alpha,\beta)
\rightarrow (ph,x,\alpha,\beta)\} \supset$ a change in
beta-time(d$\beta$) at the adjacent space $x^+$ (where $x^+=x+dx$)
such that $\{(Q,x^+,\alpha,\beta) \rightarrow
(Q,x^+,\alpha,\beta^+)\}$ where Q is the existing State value at
($x^+$,$\alpha$,$\beta$) and $\alpha,\beta$ relate to the times at
x \footnote{This rule for change in State applies in tandem with
IFE rule (i). Note also that adjacent Space entities will be
triggered for each spatial dimension - see Appendix 2.}. Since it
takes alpha-time of (pst') to cycle to the (ph) State, the
alpha-time effectively defines the speed of the IFE disturbance.

(iv) Change in beta-time(d$\beta$) of $\{(R,x,\alpha,\beta)
\rightarrow (R,x,\alpha,\beta^+)\} \supset$ a change in State(dR)
such that $\{(R,x,\alpha,\beta^+) \rightarrow
(R^+,x,\alpha,\beta^+)\}$

This algorithm defines a disturbance which moves with a constant
velocity through space and time. The disturbance has inertia and
moves indefinitely with this constant velocity - until it
interacts with another entity. The change in beta-time logically
follows the change in alpha-time.\\

\textbf{Interaction algorithm component}\\
All interactions between two IFEs are defined to occur only where
both IFEs have the same the same Space position AND Time Magnitude
$|T|$ (combined alpha-time and beta-time).

For two IFE's A $\{$($R_A,x_A,\alpha_A,\beta_A$) with
$d\alpha_A$=($s_At'$)$\}$ and B $\{$($R_B,x_B,\alpha_B,\beta_B$)
with $d\alpha_B$=($s_Bt'$)$\}$ an interaction only occurs if
$\{x_A=x_B\}$ AND $\{|T_A|=|T_B|\}$

The following check for an interaction is inserted in the
algorithm:

if \{(R,x,$|(\alpha$+d$\alpha$)+$\beta)|$) $\rightarrow$
(R',x,$|(\alpha$+d$\alpha$)+$\beta)|$\}$\supset$ INTERACTION

i.e. if there is a change in State at the current Space position
and Time magnitude \emph{of when the IFE is about to be} then an
interaction occurs. On an interaction occurring, the collapse
function is initiated (see next section).

Because of its distributed nature, no definite State or Space
position of an IFE disturbance exists until its final position is
determined by an interaction.

Further, as the starting conditions are not known then a
statistical approach must be used to calculate the probability of
interaction at a particular spatial location.\\

\textbf{Collapse algorithm component}\\
The \emph{distributed} nature of the IFE disturbance implies that
if an interaction occurs at a precisely defined combination of
State(R) and Space(x), a set of active States at Space positions
remains \emph{at the same Time Magnitude} where the specific
interaction does not occur. The collapse function removes these
components (where $\o$ indicates a State null value and the
initial State $R\neq \o$) and we define it as:

[$\{(R,x+dx,\alpha,\beta) \rightarrow (\o,x+dx,\alpha,\beta)\}$ OR
\{(R,x-dx,$\alpha$,$\beta$) $\rightarrow$
($\o$,x-dx,$\alpha$,$\beta$)\}] $\supset \{(R,x,\alpha,\beta)
\rightarrow (\o,x,\alpha,\beta)\}$

i.e. If a State IFE in a disturbance changes to a null State then
a spatially adjacent State IFE will also go to a null State. The
\emph{logical position} of this monitoring algorithm is important.
It sits in the loop which performs \emph{single (t')} increments
of alpha-time. Since this ensures continuous monitoring of
adjacent cells, and \emph{because of the precedence of logic over
temporal advance} virtually instantaneous collapses of IFE
functions can occur over over a wide region of space. It is true
to say that ``nothing moves faster than the speed of light'', but
the ``nothing'' has a reality.

The considerable debate over the process of collapse has centred
on the implication for action at a distance or for ``hidden
variables". e.g. Von Neuman (1955) asserted that for a
wave/particle its mechanism for evolution in time through space
and its mechanism for collapse are necessarily different. However,
the algorithm for collapse described above, deriving from the
precedence of logic over time
and the momentary possibility of both (R,x,$\alpha,\beta$) and (R',x,$\alpha,\beta$) negates this assertion.\\

\textbf{Spatial dimensions}\\
So far used a single spatial dimension has been used to describe
the key concepts of the theory. However, the algorithm properly
operates in 3 spatial dimensions which requires a further
refinement. This is
detailed in Appendix 2.\\

\textbf{Summary of algorithm}\\
The rules can be summarised in a logic loop:

\hspace* {-15mm} LOOP \{(R,x,$\alpha$,$\beta$) $\rightarrow$
(R',x,$\alpha$,$\beta$) \}
$\supset$ COLLAPSECHECK; ELSE LOOP\\
\hspace* {-4mm} INC \{(Q,x,$|(\alpha$+d$\alpha$)+$\beta)|$)
$\rightarrow$ (Q',x,$|(\alpha$+d$\alpha$)+$\beta)|$\}$\supset$
INTERACT

(R',x,$\alpha$,$\beta$) $\rightarrow$ (R',x,$\alpha$',$\beta$)

\{(R,x,$\alpha$,$\beta$) $\rightarrow$ (R,x,$\alpha$',$\beta$)\}
 $\supset$ \{(R,x,$\alpha$',$\beta$)$\rightarrow$
 (R',x,$\alpha$',$\beta$)\}

\{(R,x,$\alpha$,$\beta$) $\rightarrow$ (ph,x,$\alpha$,$\beta$)\}
$\supset$ \{(Q,x',$\alpha$,$\beta$) $\rightarrow$
(Q,x',$\alpha$,$\beta$')\}

\{(R,x,$\alpha$,$\beta$) $\rightarrow$ (R,x,$\alpha$,$\beta$')\}
$\supset$ \{(R,x,$\alpha$,$\beta$') $\rightarrow$
(R',x,$\alpha$,$\beta$')\}

\hspace* {-39mm} COLLAPSECHECK \{(R,x+dx,$\alpha$+s,$\beta$)
$\rightarrow$ ($\o$,x+dx,$\alpha$+s,$\beta$)\}  or
\{(R,x-dx,$\alpha$+s,$\beta$) $\rightarrow$
($\o$,x-dx,$\alpha$+s,$\beta$)\} \hspace* {10mm}$\supset$
(R,x,$\alpha$,$\beta$) $\rightarrow$ ($\o$,x,$\alpha$,$\beta$) AND
LOOP\\
\hspace* {8mm}S=S+t'\\
\hspace* {6mm} $\{S\neq st'\} \supset$ COLLAPSECHECK; ELSE S=0 AND INC \\
\hspace* {-18mm} INTERACT (R,x,$\alpha$,$\beta$) $\rightarrow$
($\o$,x,$\alpha$,$\beta$); LOOP\\

\section{Properties of the disturbance}

\hspace* {4mm} 1. A moving disturbance comprises the interrelated
fluctuating entities of State, alpha-Time, beta-Time and Space.

2. Each Space element is a distance dx apart from another: 0
indicates the position of the first element, 1 that of the next
element $\ldots$99 the $100^{th}$ element etc. Hence proceeding
from the first element to the nth element, the Space distance is
$x=ndx$.

3 A disturbance either has positive or negative movements in
Space. Thus it goes forward (x$\rightarrow$x+dx) or backward
(x$\rightarrow$x-dx)in a spatial dimension.

4 Measurement of time can only be made through change of State
\emph{- i.e. this implies that only in alpha-time can be
observed.}.

5. Each State element is h units apart from another: 0 indicates
the position of the first element, 1 that of the next (i.e.
$\{0,h,2h...\}$).\footnote{Negative States can theoretically
advance through $\{-h,-2h,-3h...\}$. However, they will not be
discussed in this paper}

6. A change in State entity dR triggers a change in alpha-time
(d$\alpha$=st' where s represents the Time$\rightarrow$State
trigger link such that after s cycles of t' an advance in State dR
is triggered). Thus the Time recorded by a particle to reach a
State$\rightarrow$Space trigger point of ph is $pst'$.

If in a time T a disturbance with a State$\rightarrow$Space
trigger of ph advances by r changes in State at the final Space
position then the total time is T = (npst'+rst'+nt$^*$).

7. All spatially local measurements of time are performed through
changes in State phenomena. Each IFE disturbance can therefore
\textit{\textbf{itself measure time only through a change in
alpha-time}}. Each change of State triggers a local change in
alpha-time where local time applies to the Space position of the
disturbance.

8. It is notable that the (unresolved) total Time
T=($\alpha$,$\beta$) can be represented as a complex number. Using
a notation of beta-time as real and alpha-time as imaginary: \beq
\b{T} = n t^* + \imath (n p + r) st' \eeq or where $z =( p +
r/n)s$ : \beq \b{T} = n (t^* + \imath z t') \eeq

9.  All interactions occur at the same Time Magnitude $ {\mid}  \b
{T}  {\mid} = \sqrt{ {T}{T^*}} = \sqrt{(nt^{*})^{2} +
(npst'+rst')^{2}}$. For large n the residual rst' alpha-time
component (i.e. the additional State changes at a spatial
position) in calculations of time magnitude can often be ignored.
For increasingly small distances, however, the rst' component
assumes an increasing proportion of the total Time.

10. Frequency is defined as f = $\frac{1}{\imath (st')}$. The
(st') term indicates the Time to move from one State position to
another.

11. Speed is defined as the rate of change of Space over Time.
\beq v = \frac{ndx}{|T|} = \frac{ndx}{\sqrt{(nt^*)^2+(npst'+
rst')^2}} \eeq

12. A maximum speed is implied at which an disturbance can
propagate through the Space medium. This occurs when the
State$\rightarrow$Space trigger point p is zero. i.e.

\beq v_{max} = \frac{dx}{\sqrt{(t^*)^2+(0 + \frac{rst'}{n})^2}}
\approx \frac{dx}{t^*} \eeq

The denominator represents\footnote{over a reasonable (any
measurable) distance: $n\gg(rst')$} the time taken to move a
single spatial distance by an entity with no State changes
occurring. $v_{max}=c$ is the speed of light. The constant c
consequently connects the smallest possible change in spatial
distance dx to the smallest discrete increase of beta-time
$t^{*}$. The absence of alpha-time in the time magnitude explains
why such a speed cannot be exceeded. For an entity (such as a
photon) travelling at this speed, no time is experienced by that
entity (experienced time $=$ alpha-time $=npst'$).

13.  Each change in Space triggers a change in beta-time -
effectively the Time for the IFE disturbance to propagate to an
adjacent Space position. Since there is empirically a fine
gradation in possible speeds, then $t^* > t'$ and generally $pst'
\gg t^*$ \footnote{If there were a change in speed from
$c=\frac{dx}{\sqrt{(t^*)^2+0}}$ to the next fastest speed $c' =
\frac{dx}{\sqrt{(t^*)^2 + (spt')^2}}$ (and setting p=s=1) then
were $t'=t^*$ then $c'\sim \frac{d}{\sqrt{2(t^*)^2}}
=\frac{c}{\sqrt{2}}$ which is not the case. Hence $t'\ll t^*$}

14. Wavelength $ \lambda =   v/f = \imath st'v$

\beq  \lambda = \frac{\imath(dx)(st')}{\sqrt{(t^*)^2+(pst')^2}}
\eeq

15. Energy is defined as the rate of change of State. If measured
at a constant spatial position the this will be the rate of change
of State as measured in alpha-time (measurement of energy by a
\emph{moving} disturbance is covered in a later section). Then for
a single change in state: $e= \frac{h}{(st')}$. h (Planck's
constant) represents the smallest possible discrete increase in
State. In a collision of two IFEs A and B with state transition
times of $s_A$ and $s_B$ (i.e. where energies are
$\frac{h}{s_At'}$ and $\frac{h}{s_Bt'}$) then in a time $s_A$ A
moves h State units and in a time $s_As_B$ A moves $s_Bh$ units;
correspondingly for B in a time $s_As_B$ B moves $s_Ah$ units.
Thus in a time $s_As_B$ there is a total State change of
$h(s_A+s_B)$ units. Therefore the total combined energy is:
$e_{tot}=\frac{h(s_A + s_B)}{s_As_B}$

16.  The disturbance's spatial identity bifurcates at the point of
the State$\rightarrow$Space trigger. This encompasses \emph{both}
a change in alpha-time at the existing Space and a change in
beta-time at the adjacent space. An ambiguity results: both where
an entity is located in Space and what its State is are undefined.
For a \emph{given} time magnitude $|T|$ this ambiguity can be
captured through $\sqrt{(nt^*)^2+(npst' + rst')^2}=|T|$. Since n
and r are variables, different combinations of State and Space
positions can form the same Time magnitude $|T|$ from variable
alpha-time and beta-time constituents. A fixed $|T|$ of magnitude
$|rst'|$ forms a ``temporal arc". This is easily represented for a
null State$\rightarrow$Space trigger (e.g. a
photon) where p=0 (see Diagram 2 below).\\

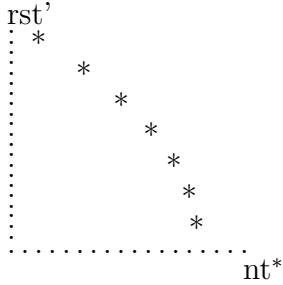
DIAGRAM 2 - temporal arc for a photon at time magnitude $|rst'|$ \\
\hspace* {-6mm} \\rst'
\\[-1.6mm]
 \vdots\hspace* {2mm}*
\\[-1.6mm]
 \vdots\hspace* {8mm}*
\\[-1.6mm]
\vdots\hspace* {13mm}*
\\[-1.6mm]
\vdots\hspace* {17mm}*
\\[-1.6mm]
\vdots\hspace* {20mm}*
\\[-1.6mm]
\vdots\hspace* {22mm}*
\\[-1.6mm]
\vdots\hspace* {23mm}*
\\[-3.2mm]
\ldots\ldots\ldots\ldots\ldots\ldots \hspace* {23mm}
\\[-1.6mm]
 \hspace* {30mm}
nt$^{*}$

\textbf{All points on the temporal arc have the same time
magnitude.}\\

\section{Time Magnitude (over large distances)}

The time \textit{measured/experienced} (alpha-time) by a spatially
moving disturbance can be compared with that of a State-changing
but spatially stationary one.

From the example outlined earlier in Diagram 1, the time measured
by a disturbance moving in Row A from point 1 to point n in the
same row can be compared with the time difference measured between
stationary entities in row B at spatial points 1 and n.

Since all interactions occur at the same time magnitude then at
the point of interaction at the nth spatial position, the State
IFEs in both rows have the same time \textit{magnitude}.

For the spatially moving disturbance, the time
\textit{experienced} $A_\alpha=pst'$ is simply the alpha-time
$npst'$. However, because it moves spatially, then from equation
(1) and assuming n is large, its time magnitude $|T|$ comprises
both alpha-time and beta-time: $|T|=n\sqrt{(t^*)^2 + (pst')^2}$.
The spatially stationary disturbance in the second row interacts
at the same time magnitude of $|T|=n\sqrt{(t^*)^2 + (pst')^2}$. It
therefore experiences \emph{alpha-time} of $B_\alpha =
|n\sqrt{(t^*)^2 + (pst')^2}|$.

Differences in experienced time between moving and stationary
entities all stem from the indirect addition of beta-time. Thus
$B_\alpha<A_\alpha$. For this simple reason ``moving clocks run
slow". This can be calculated formally.\footnote{The probability
of an interaction at a specific spatial point will decrease with
distance as the larger the arc the greater the probability of an
interaction elsewhere on the circumference of the arc. Thus for a
beam of photons, we would expect the intensity of the beam to
diminish - without the energy of an individual photon being
weakened.} The total amount of time taken by the moving
disturbance in row A to move from position 1 to position n (where
for convenience $z = (p + \frac{r}{n})s$) is $T = n(t^*+ \imath
zt')$. The magnitude is:

\beq |T |= n \sqrt{(t^*)^2 +(zt')^2} \eeq

This simple equation entirely captures the theory of special
relativity, for $|T|$ expresses the total time magnitude and (zt')
represents the time ``experienced" by the moving IFE. To
demonstrate accordance with the familiar Lorentz/Einstein model:

\beq \textmd{Speed } v = \frac{nd}{n \sqrt{(t^*)^2 +(zt')^2}} =
\frac{d}{\sqrt{(t^*)^2 +(zt')^2}} \eeq

For the photon travelling over a significant distance  there is no
State$\rightarrow$Space trigger point (i.e. p=0) and r/n is very
small compared with $t^*$. Then: \beq \textmd{Speed } c =
\frac{nd}{nt^*} = \frac{d}{t^*} \eeq

Rearranging  (7):

$$ |T| = n\left( \frac{(t^*)^2}{\sqrt{(t^*)^2 +(zt')^2}} +
\frac{(zt')^2}{\sqrt{(t^*)^2 +(zt')^2}}\right) $$

Substituting from (8) and (9) into the first part of the
expression and rearranging the second part:

$$ |T| = \frac{nv(t^*)}{c} + n(t^*
zt')\frac{(zt')}{t^*}\sqrt{\frac{1}{(t^*)^2 + (zt')^2}} $$

Further rearranging:

$$ |T| = \frac{nv(t^*)}{c} + n(t^* zt')\sqrt{\frac{(t^*)^2 +
(zt')^2 - (t^*)^2}{(t^*)^2[(t^*)^2+(zt')^2]}} $$

From which we obtain:

\beq |T| = \frac{nv(t^*)}{c} + n(t^* zt')\sqrt{\frac{1}{(t^*)^2} -
\frac{1}{(t^*)^2 + (zt')^2}} \eeq

But from (8) and (9) we have:

\beq \frac{\sqrt{c^2 - v^2}}{c} = t^* \sqrt{\frac{1}{(t^*)^2} -
\frac{1}{(t^*)^2 + (zt')^2}} \eeq

Substituting this expression into (10) we obtain:

\beq\ |T| = \frac{nv(t^*)}{c} + n \frac{\sqrt{c^2 - v^2}}{c} (zt')
\eeq

Now in terms of distance travelled x:
$$x =  c(nt^*)$$

Substituting into (12) we arrive at:

$$ |T| = n(zt')\sqrt{1-v^2 / c^2} + (v / c^2)x $$

Since n(zt') corresponds to $\tau$ the amount of time experienced
from the perspective of the moving entity (often referred to as
the proper time) and $|T|$ corresponds to the time observed by a
stationary observer, this is the familiar Einstein-Lorentz
expression:

\beq\tau = \gamma(|T | -  (v x/ c^2 ))   \textnormal{   where }
\gamma = (1 - v^2 /c ^2)^{-1/2} \eeq

The simplicity and explanatory power of this approach in equation
(7) is notable by comparison.

All ``relativistic" effects are fundamentally underpinned by time
and time alone. Apparent alterations in distance arise from the
perception of measured space through velocities which ultimately
relate to differences in experienced time derived from combination
of beta-time and alpha-time.

\section{Energy viewed by a moving disturbance}

If a matter source disturbance A is stationary at the origin and a
matter observer disturbance B, starting from spatial position
$x_0$, moves \emph{away from} the source with a speed which it
measures as $\frac{dx}{p_2s_2t'}$ then the observer will infer the
rate of change of State of the source \emph{through changes in
State directed to the observer by photons i.e. disturbances which
move at the speed of light}. The apparent rate of change of State
of the source will therefore depend on both the ``intrinsic" rate
of change of State of the source and the apparent speed of
movement between the source and the observer.

To aid calculation, a time interval can be deliberately selected
based on the speed of movement of the \textit{observer}:
$T_0=pst'$ (where p is the State$\rightarrow$Space trigger). The
first State change of the source is noted by the moving observer
at spatial position $x_1$ and the last State change of the source
at the end of this interval is observed by the moving disturbance
at spatial position $x_2$.

Because the interval of time $pst'$ is measured \emph{by the
moving observer disturbance} which moves at speed which it
perceives as $\frac{dx}{pst'}$, this implies that one ``skip'' of
$t^*$ will occur during this time interval which will not be
experienced by the observer.

Each change of State will relay via a photon from the source to
the observer at the speed of light $c=\frac{dx}{t^*}$. $x_1$
occurs at a coincidence (same time magnitude and spatial position)
of the observer and the first photon from the first State position
from the source. $x_2$ occurs at a coincidence between the
observer and a photon emitted from the source after a
source-measured time interval of $pst'$. For $x_1$ we have:

\beq x_1=\frac{dx}{t^*}t_1=x_0+\frac{dxt_1}{pst'} \eeq (if the
observer was moving \textit{towards} the source then
$x_1=x_0-\frac{dxt_1}{pst'})$

There will be no spatial movement of the electromagnetic
disturbance during the time period spent entirely on State
movements by the source at the same fixed Space position.

However, an additional skip of beta-time in the observer has to be
accounted for after an interval of $pst'$ during which the photon
will move. This effectively adds an extra distance of $dx = ct^*$
onto the distance travelled by the photon during the time $pst'$
measured by the observer.

The point of coincidence between the photon and the observer
occurs when both photon and observer have the same time magnitude
and spatial position. Thus the apparent time \emph{as measured by
the observer}, taken for the source's State movement is
\emph{shortened to} $\sqrt{(pst')^2 - (t^*)^2}$ (which equates to
$pst'$ as measured by the source)

\beq x_2=\frac{dx(t_2 +  \sqrt{(pst')^2 - (t^*)^2})}{t^*}
 = x_0+\frac{dxt_2}{pst'} \eeq

Then from (14) and (15) $t_1=\frac{x_0t^* pst'}{dx(pst'-t^*)}$ and
$t_2=\frac{t^* pst' (x_0 - dx\sqrt{(pst')^2 - (t^*)^2}
)}{dx(pst'-t^*)}$

$t_1-t_2= (pst') \frac{ \sqrt{(pst')^2 - (t^*)^2}}{pst'-t^*}$

The original time interval of $pst'$ represents the period $T_0$
from the perspective of the unmoving source at the origin. The
apparent period from the perspective of the moving disturbance
will be $T' = t_2-t_1$

i.e. $T'=T_0 \frac{\sqrt{(pst')^2 - (t^*)^2}}{pst'-t^*}$

And (where $e=\frac{h}{st'}$) the apparent energy $e'= e_0
\frac{pst'-t^* }{\sqrt{pst')^2 - (t^*)^2}}$

Using $\frac{v}{c} = \frac{t^*}{pst'}$ then $\frac{pst'-t^*
}{\sqrt{(pst')^2 - (t^*)^2}} = \sqrt{\frac{(pst'-t^*)^2}{(pst')^2
- (t^*)^2}} = \sqrt{\frac{pst'-t^*}{pst' + t^*}} =
\sqrt{\frac{1-\frac{v}{c}}{1+\frac{v}{c}}}$

Thus $e'=e_0 \sqrt{\frac{1-\frac{v}{c}}{1+\frac{v}{c}}}$.

For an observer moving towards the source, this would be: $e'=e_0
\sqrt{\frac{1+\frac{v}{c}}{1-\frac{v}{c}}}$

\section {Three spatial dimensions}

Empirically the \emph{speed} of a photon is isotropic in all
directions. An immediate challenge arises from the
\emph{distributed} nature of the graph model.

A difference in speed would appear to arise between measurements
taken in different coordinate systems. A distance dx measured
along the x-axis would take time $t^*$, providing a speed
$c=\frac{dx}{t^*}$. However, if the disturbance is measured moving
in more than one spatial dimension, e.g. along the diagonal of a
cube formed across x, y and z axes then the distance travelled is
$\sqrt{3}dx$. Were the total time to be a sum of the three times
$t^*$ (i.e. movement occurs in a time $3t^*$) this would create a
variation in the speed of the photon disturbance. There would
equally be a discrepancy if the total time taken is $t^*$. Yet
empirically the speed of a photon is constant and independent of
the direction of travel.

The approach suggested by this paper is of multiple dimensions of
time. If we assume that each dimension of space is associated with
a separate dimension of time, then since photons have no
alpha-time changes for changes in Space, allocating the
\emph{beta}-times across axes $(Re, \imath, \jmath)$ as $t^*,
\imath t^*, \jmath t^*$ the total time for a photon moving one
spatial position on each spatial axis is $T=t^* + \imath t^* +
\jmath t^*$. This provides a time magnitude (where for a time
magnitude $|A|_{\imath \jmath}$ we calculate first the $\imath$
component and then the $\jmath$ component):

$|T|_{ij}=|t^* + \imath t^* + \jmath t^*|_{ij}=||t^* + \imath
t^*|_\imath + |\jmath t^*|_\imath |_j = |\sqrt{2(t^*)^2} + \jmath
t^*|_\jmath = \sqrt{3}t^*$

For a distributed photon disturbance moving across the x and y and
z axes,  the disturbance might be found to be located (through an
interaction) as having moved one spatial position \emph{on the x
axis only}. In this case the speed = $\frac{dx}{t^*}$. The
disturbance might also be found to be located - again through an
interaction - having moved across the x, y and z axes, in which
case the speed = $\frac{\sqrt{3}dx}{\sqrt{3}t^*}=\frac{dx}{t^*}$.
Thus the three dimensions of time map neatly onto the three
spatial dimensions and isotropy of speed in Space-Time has been
preserved.

However, for a disturbance other than a photon, State movements
are implicit in Space movements and alpha-time advances
necessarily occur in even a single movement across Space.

If we continue to limit to 3 time dimensions, then these
alpha-time components cannot be simply allocated to the $\imath$
axis only. Were this to be the case, we would obtain: $t_x=
t^*+\imath s_xt'$, $t_y=\imath t^*+\imath s_yt'$, $t_z= \jmath
t^*+\imath s_zt'$ and inevitable interference would occur between
the alpha-time and beta-time components.

However, if the number of time dimensions is kept to three, an
elegant allocation mechanism can preserve isotropy of speed for a
moving particle across three Space dimensions.

The alpha-time and beta-time components can combine in different
ways on the axes. The \emph{logical} ordering of sequences of
combinations will therefore manifest accordingly.

The advance in space can be viewed as a ``diagonal'' progress of
$|dx+dy+dz|$ and the advance in time as $|t_x+t_y+t_z|$. We can
also view there being separate components that logically and
temporally follow one-another. Thus $|dx|+|dy|+|dz|$ takes an
amount of time $|t_x|+|t_y|+|t_z|$. The speeds measured according
to either of these two methods \emph{must be the same}.

Thus
$\frac{|dx+dy+dz|}{|t_x+t_y+t_z|}=\frac{|dx|+|dy|+|dz|}{|t_x|+|t_y|+|t_z|}$

 i.e. $|t_x+t_y+t_z|=\frac{\sqrt{3}}{3} \{|t_x|+|t_y|+|t_z|\} =
\frac{1}{\sqrt{3}}\{|t_x|+|t_y|+|t_z|\}$

Combination requires us to consider the 12 possible alternative
formulations for \textit{logical} combination of components. Since
there can only be 2 axes that combine, followed by a further
combination and then a next, thus, considering x, y and z, we have
in logical order:

first $t_x$ then $t_y$, or first $t_y$ then $t_x$

$\{(x+y)+z\}, \{(y+x)+z\}, \{(x+z)+y\}, \{(z+x)+y\}, \{(y+z)+x\},
\{(z+y)+x\}$

and in addition:

$\{z+(x+y)\}, \{z+(y+x)\},\{y+ (x+z)\}, \{y+ (z+x)\},\{x+
(y+z)\},\{x+(z+y)\}$

This requires careful co-ordination of the different time
components. A movement in x and y, can be \emph{compensated} for
by a movement in the z time contribution. This requires that we
divide the time movement for a single movement in space into
\textit{two} logical components of time increase i.e.
$T_x=t_{x1}+t_{x2}; T_y=t_{y1}+t_{y2}; T_z=t_{z1}+t_{z2}$

The beta-time components for $t_x, t_y, t_z$ will be along
different axes $Re, \imath,\jmath$. Solution of the combination of
these components can be done using:\\

$\hspace{-8mm}$ $t_{x1} = t^* +  \jmath s_{x1}t'$ ; $t_{x2} =
\jmath t^* + s_{x2}t'$\\
$ t_{y1} = - \jmath t^* + \imath \jmath s_{y1}t'$ ;
$t_{y2} = - \imath \jmath t^* + \jmath s_{y2}t'$ \\
$t_{z1} = \imath t^* - s_{z1}t'$ ;
$t_{z2} = t^* - \imath s_{z2}t'$\\

and through:

(i) isotropy of time: $|T_x|=|T_y|=|T_z|$

and selection of:

(ii) $s_{x1}t't^* + s_{z1}t't^* + s_{x2}t' s_{z1}t' = s_{x2}t't^*
+ s_{y2}t't^* + s_{x1}t' s_{y2}t' $\\

The time magnitudes of each direction component are:

$\hspace{-10mm}$ $|T_x|=|t_{x1}+t_{x2}|_{\imath\jmath}=
|(t^*+s_{x2}t')+\jmath (t^* + s_{x1}t')|_\jmath
 = \sqrt{2(t^*)^2 + (s_{x1}t')^2 + (s_{x2}t')^2 + 2s_{x1}t't^* + 2s_{x2}t't^*}$

$\hspace{-10mm}$ $|T_y| = |t_{y1}+t_{y2}|_{\imath\jmath}=
\sqrt{2(t^*)^2 + (s_{y1}t')^2 + (s_{y2}t')^2 - 2s_{y1}t't^* -
2s_{y12}t't^*}$

$\hspace{-10mm}$$|T_z| = |t_{z1}+t_{z2}|_{\imath\jmath} =
\sqrt{2(t^*)^2 + (s_{z1}t')^2 +
(s_{z2}t')^2 - 2s_{z1}t't^* - 2s_{z2}t't^*}  $\\

The overall time magnitude is:

$|T| = | (2t^* + s_{x2}t' -s_{z1}t')  + \imath(t^* - s_{z2}t') +
\jmath(s_{x1}t' + s_{y2} +  \imath(-t^* + s_{y1}t')\}|_{\imath\jmath}$\\

$\hspace{-15mm}$ Thus $|T| =| [(2t^* + s_{x2}t' -s_{z1}t' )^2 +
(t^* - s_{z2}t')^2]^\frac{1}{2} + \jmath[(s_{x1}t' + s_{y2}t')^2 +(-t^* + s_{y1}t')^2]^\frac{1}{2}|_\jmath$\\

i.e. $|T| = [(s_{x1}t')^2 +(s_{x2}t')^2 +(s_{y1}t')^2
+(s_{y2}t')^2+(s_{z1}t')^2 +(s_{z2}t')^2 +6(t^*)^2 +4s_{x2}t't^* -
4s_{z1}t't^* - 2s_{z2}t't^* -2s_{y1}t't^* - 2s_{x2}t's_{z1}t' +
2s_{x1}t's_{y2}t']^\frac{1}{2} $\\

Using the expansions above, we have
$(|T_x|+|T_y|+|T_z|)^2=(s_{x1}t')^2 +(s_{x2}t')^2 +(s_{y1}t')^2
+(s_{y2}t')^2+(s_{z1}t')^2 +(s_{z2}t')^2 +6(t^*)^2  + 2s_{x1}t't^*
+ 2s_{x2}t't^* + 2s_{y1}t't^* + 2s_{y2}t't^* + 2s_{z1}t't^* +
2s_{z2}t't^* + 2|T_x||T_y| + 2|T_x||T_z| +
2|T_y||T_z|$\\

The multiples $2|T_x||T_y|$, $2|T_x||T_z|$ and $2|T_y||T_z|$ can
be calculated:

Since $|T_x|=|T_y|=|T_z|$ we have:

$(s_{z1}t')^2+(s_{z2}t')^2=(s_{y1}t')^2+(s_{y2}t')^2 -
2s_{y1}t't^* - 2s_{y2}t't^* - 2s_{z1}t't^* - 2s_{z2}t't^*$

$\sqrt{(s_{z1}t')^2+(s_{z2}t')^2 +2(t^*)^2} = \sqrt{|T_y|^2 + 2t^*
(s_{z1}t'+s_{z2}t')} = \sqrt{|T_x|^2 + 2t^* (s_{z1}t'+s_{z2}t')}$

$\hspace{-30mm}  (s_{z1}t')^2+(s_{z2}t')^2 +2(t^*)^2 =
\sqrt{|T_x|^2 |T_y|^2 + |T_y|^2 2t^* (s_{z1}t'+s_{z2}t') + |T_x|^2
2t^* (s_{z1}t'+s_{z2}t') + 4t^* (s_{z1}t'+s_{z2}t')^2}$

$ = \sqrt{(|T_x| |T_y|)^2  +(2t^*(s_{z1}t'+s_{z2}t'))^2 + 2t^*
(s_{z1}t'+s_{z2}t')(|T_x|^2 + |T_y|^2)}$

$ = \sqrt{|T_z|^4  +(2t^*(s_{z1}t'+s_{z2}t'))^2 + 4t^*
(s_{z1}t'+s_{z2}t')(|T_z|)^2}$

$ = \sqrt{(|T_z|^2  -2t^*(s_{z1}t'+s_{z2}t'))^2 } = |T_z|^2 +
2t^*(s_{z1}t'+s_{z2}t')$

$ = |T_x||T_y| + 2t^*(s_{z1}t'+s_{z2}t')$

i.e. $|T_x||T_y|=(s_{z1}t')^2+(s_{z2}t')^2 -
2t^*(s_{z1}t'+s_{z2}t') + 2(t^*)^2$

Likewise $|T_x||T_z|=(s_{y1}t')^2+(s_{y2}t')^2 -
2t^*(s_{y1}t'+s_{y2}t') + 2(t^*)^2$

And $|T_y||T_z|=(s_{x1}t')^2+(s_{x2}t')^2 +
2t^*(s_{x1}t'+s_{x2}t') +
2(t^*)^2$\\

Thus $(|T_x|+|T_y|+|T_z|)^2= 3\{(s_{x1}t')^2 +(s_{x2}t')^2
+(s_{y1}t')^2 +(s_{y2}t')^2+(s_{z1}t')^2 +(s_{z2}t')^2 +6(t^*)^2 +
2s_{x1}t't^* + 2s_{x2}t't^*  - 2s_{y1}t't^*  -2s_{y2}t't^* -
2s_{z1}t't^* - 2s_{z2}t't^* \}$

Using the earlier expression for $ |T|$ and (ii):

$ |T|= [(s_{x1}t')^2 +(s_{x2}t')^2 +(s_{y1}t')^2
+(s_{y2}t')^2+(s_{z1}t')^2 +(s_{z2}t')^2 +6(t^*)^2 +2s_{x1}t't^*
+2s_{x2}t't^* -2s_{y1}t't^* - 2s_{y2}t't^*
    - 2s_{z1}t't^* - 2s_{z2}t't^*]^\frac{1}{2}$

So $(|T_x|+|T_y|+|T_z|)^2=3|T_x+T_y+T_z|^2$

i.e. $|T_x|+|T_y|+|T_z| = \sqrt{3}|T_x+T_y+T_z|$

Which accords with a constant speed independent of the direction
of movement. Thus:

$\frac{\sqrt{3}dx}{|T_x+T_y+T_z|}=\frac{3dx}{|T_x|+|T_y|+|T_z|}$

And isotropy of Space is preserved.

\section {Speculation on Gravitation}

Analysis has focused on \emph{changes} in time magnitudes and the
temporal arc formed by such intervals. However, the \emph{total}
time of a disturbance should additionally be considered.

Given that the age of the universe is estimated at at least ten
billion years, the total alpha-time \emph{of the measurable matter
of the universe around us} is pretty much a constant for
measurements completed in the last hundred years. This follows
because firstly experiments in our purview of a hundred years will
not have any significant impact on the total time magnitude of an
object. Secondly the (``heavy'') objects around us do not move at
speeds close to the speed of light, and therefore the alpha-times
will be comparatively close to the total time magnitudes.

The focus that is required is on alpha-time. Consider that changes
have already been made for some time by graviton emission -
\emph{i.e. temporal arc is already formed}. It is assumed that,
just as for the variable State positions highlighted earlier,
there is an equivalent for the total alpha-time of a disturbance
which can be distributed across a temporal arc of the entirety of
the alpha-time. A range of possible States will therefore be
distributed across a temporal arc \emph{of the entirety of the
alpha-time}. A key assumption is that the States of one
disturbance can impact on the States of another disturbance.

Consider two disturbances a distance r apart: an observational
photon disturbance A with energy $\frac{h}{s_At'}$ and total
alpha-time $T_A$ and a slow-moving source disturbance B with
energy $\frac{h}{s_Bt'}$ and a total alpha-time $T_B$.

The initial State of A at the first Spatial position is due to A's
initial State and the State contribution of B.

To calculate this we adjust measures of Space through State
changes calibrated in $s_Bt'$:

i.e. initial State = $\frac{T_A h}{s_At'} +
(\sqrt{(\frac{T_B}{s_Bt'})^2-(\frac{rt^*}{dx s_Bt'})^2} )h $

The States of the two disturbances will, in time $(s_At')$ as
measured at A, have advanced State \emph{at B} by
$\frac{s_At'h}{s_Bt'}$ and at A by h.

Later State =  $\frac{T_A h}{s_At'} + h +
(\sqrt{(\frac{T_B}{s_Bt'} + \frac{s_At'}{s_Bt'})^2-(\frac{rt^*}{dx
s_Bt'})^2}) h $

i.e. rate of change of State of the photon in the first Space
position $e_0 = \frac{h + (\sqrt{(\frac{T_B}{s_Bt'} +
\frac{s_At'}{s_Bt'})^2-(\frac{rt^*}{dx s_Bt'})^2}) h -
(\sqrt{(\frac{T_B}{s_Bt'})^2-(\frac{r
\frac{t^*}{dx}}{s_Bt'})^2})h}{s_At'} $

To calculate the change of State of the photon at the adjacent
spatial position (i.e. which is a distance dx closer to the
source):

Initial State = $\frac{T_A h}{s_At'} +
(\sqrt{(\frac{T_B}{s_Bt'})^2-(\frac{r\frac{t^*}{dx}-t^*}{s_Bt'})^2}
)h $

Later State =  $\frac{T_A h}{s_At'} + h +
(\sqrt{(\frac{T_B}{s_Bt'} +
\frac{s_At'}{s_Bt'})^2-(\frac{r\frac{t^*}{dx}-t^*}{s_Bt')})^2}) h
$

i.e. rate of change of State of the photon in the second Space
position $e_1 =  \frac{h + (\sqrt{(\frac{T_B}{s_Bt'} +
\frac{s_At'}{s_Bt'})^2-(\frac{r\frac{t^*}{dx}-t^*}{s_Bt'})^2}) h -
(\sqrt{(\frac{T_B}{s_Bt'})^2-(\frac{r
\frac{t^*}{dx}-t^*}{s_Bt'})^2})h}{s_At'}$

i.e. the difference in energy for the photon between the first and
second spatial positions is $e_1-e_0$:

\hspace{-12mm}$e_1-e_0=\frac{(\sqrt{(\frac{T_B}{s_Bt'}+\frac{s_At'}{s_Bt'})^2-(\frac{r\frac{t^*}{dx}-t^*}{s_Bt'})^2})h-(\sqrt{(\frac{T_B}{s_Bt'})^2-(\frac{r\frac{t^*}{dx}-t^*}{s_Bt'})^2})h
-(\sqrt{(\frac{T_B}{s_Bt'}+\frac{s_At'}{s_Bt'})^2-(\frac{r\frac{t^*}{dx}}{s_Bt'})^2})h-(\sqrt{(\frac{T_B}{s_Bt'})^2-(\frac{r\frac{t^*}{dx}}{s_Bt'})^2})h}{s_At'}$

\hspace{-18mm}$= \frac{h}{s_at'} [(\frac{r \frac {t^*}{dx} -
t^*}{s_Bt'})
\{\sqrt{\frac{(\frac{T_B}{S_Bt'}+\frac{s_At'}{s_Bt'})^2}{(\frac{\frac{rdx}{dt^*}-t^*}{s_Bt'})^2}-1}
-
\sqrt{\frac{(\frac{T_B}{S_Bt'})^2}{(\frac{\frac{rdx}{dt^*}-t^*}{s_Bt'})^2}-1}
\} -(\frac{r \frac
{t^*}{dx}}{s_Bt'})\{\sqrt{\frac{(\frac{T_B}{S_Bt'}+\frac{s_At'}{s_Bt'})^2}{(\frac{\frac{rdx}{dt^*}}{s_Bt'})^2}-1}
-
\sqrt{\frac{(\frac{T_B}{S_Bt'})^2}{(\frac{\frac{rdx}{dt^*}}{s_Bt'})^2}-1}
\}]$

Using a Binomial expansion:

$\sim \frac{h}{2s_at'} [(\frac{r \frac {t^*}{dx} - t^*}{s_Bt'})
\{\frac{(\frac{T_B}{S_Bt'}+\frac{s_At'}{s_Bt'})^2}{(\frac{\frac{rdx}{dt^*}-t^*}{s_Bt'})^2}
-
\frac{(\frac{T_B}{S_Bt'})^2}{(\frac{\frac{rdx}{dt^*}-t^*}{s_Bt'})^2}
\} -(\frac{r \frac
{t^*}{dx}}{s_Bt'})\{\frac{(\frac{T_B}{S_Bt'}+\frac{s_At'}{s_Bt'})^2}{(\frac{\frac{rdx}{dt^*}}{s_Bt'})^2}
- \frac{(\frac{T_B}{S_Bt'})^2}{(\frac{\frac{rdx}{dt^*}}{s_Bt'})^2}
\}]$

$=\frac{h}{2s_At'}[(\frac{s_Bt'}{r\frac{t^*}{dx}-t^*})\{
(\frac{T_B}{s_Bt'}+\frac{S_At'}{S_Bt'})^2  -(\frac{T_B}{s_Bt'})^2
   \}-(\frac{s_Bt'}{r\frac{t^*}{dx}})\{(\frac{T_B}{s_Bt'}+\frac{s_At'}{s_Bt'})^2-(\frac{T_B}{s_Bt'})^2\}]$

$=\frac{h}{2s_At'}
(\frac{s_Bt'}{r\frac{t^*}{dx}-t^*}-\frac{s_Bt'}{r\frac{t^*}{dx}})
\{(\frac{T_B}{s_Bt'}+\frac{s_At'}{s_Bt'})^2-(\frac{T_B}{s_Bt'})^2\}$

$\sim \frac{h t^*s_Bt'}{2s_At' r^2}(2\frac{T_b}{s_Bt'}
\frac{s_At'}{s_Bt'}+(\frac{s_At'}{s_Bt'})^2)$ assuming that r is
very large c.f. dx

Assuming that $T_B \gg s_At'$ (where $T_B$ is the age of the
source disturbance and $s_At'$ is the time taken for a single
State change of the photon) then the change in energy of the
photon:

\beq e_1-e_0\sim \frac{hcT_B dx}{r^2 (s_At') (s_Bt')} \eeq

This energy change occurs in a time $s_At'$. Thus the rate of
change of energy $= \frac{2hcT_B dx}{r^2s_A t' s_Bt'}$

For a photon, since velocity is constant = c, change in energy
relates to change in mass and Force $F = \frac{d(mv)}{dt} = c
\frac{dm}{dt}$. Since $m=\frac{e}{c^2}$ then $F =
\frac{1}{c}\frac{de}{dt}$ and:

\beq F =\frac{hT_B dx}{r^2 (s_At')( s_Bt')} \eeq

But since $m_A = \frac{h}{s_At'c^2}$ and $m_B =
\frac{h}{s_Bt'c^2}$ then $F = \frac{G m_A m_B}{r^2}$

Thus as $h, c, T_B, dx$ are all constants, this implies

\beq G= \frac{T_B dxc^4}{h} \eeq

dx therefore differs from the Planck distance. Assuming Planck's
constant $h = 6.63X10^{-34}$ Js, speed of light $c=3X10^8 m/s$,
the gravitational constant $G=6.67X10^{-11}$ and the age of the
universe $T_B$ as approximately 10 billion years
($=3.15X10^{17}s$) then: $ dx = \frac{Gh}{T_B c^4} \sim
1.73X10^{-95}m$ which is \emph{much} smaller than the Planck
distance.

Note that this calculation is for a change in energy for a
\emph{single} Space position movement dx. For a larger change in
spatial position ndx, the calculation is considerably more complex
as the changes in energy have to be \emph{accumulated} across each
spatial position and then reflected back into the calculation for
the influence of $T_B+d(s_At')$.

This calculation implies that G varies over time and is
\emph{increasing}. Additionally the gravitational force exerted by
a disturbance that has been moving very fast over a long time
period will be \emph{lower} than that for a slower-moving one. The
challenge is that we do not have the opportunity to measure
gravitational forces produced by disturbances that have been
moving very fast for a very long time as they tend to be extremely
low in mass.

\section{Conclusions}

The multi-dimensional time approach underpins significant aspects
of the theories of relativity and quantum physics - including why
the speed of light has a maximum, perceived differences in
experienced time for moving and stationary entities, how the
concepts for the speed of light c and Planck's constant h are
derived more fundamentally from the units of alpha-time and
beta-time and non-localised effects involving the collapse
function.

A further paper describes the statistical consequences of defined
interaction at a specified Time Magnitude and the bifurcation of
identity at the point of a change in Space. Computer models and
discussion are available from the author on request.

\section {Acknowledgements}

The assistance of University College London Computer Science
department is gratefully acknowledged.

\section{References}

\hspace* {5mm}

(1) Feynman R (1982) "Simulating Physics with
Computers",International Journal of Theoretical Physics,
21:467-488,

(2) Jeffreys H (1939): ``Theory of Probability" , OUP.

(3) Gauss C F ``Werke" (1863-1933), Gottingen: Gesellschaft der
Wissenschafften,

(4) Jaroszkiewicz G (2000) "Causal Implication and the Origin of
Time Dilation" General Relativity and Quantum Cosmology, abstract
gr-qc/0008022

(5) Minsky M (1982) ``Cellular vacuum", International Journal of
Theoretical Physics, 21(6/7):537-551

(6) Olds C D, Lax A, Davidoff G (2000) ``The Geometry of Numbers",
The Mathematical Association of America

(7) Pearl J (2000) ``Models, Reasoning and Inference", Cambridge
University Press

(8) Tegmark (1997) ``On the dimensionality of spacetime",
Classical and Quantum Gravity, 14, L69-L75

(9) Von Neumann J (1955) ``Mathematical Foundations of Quantum
Mechanics", translated by R.T. Berger, Princeton. Princeton
University Press.

\end{document}